# Business Process Mining

Asef Pourmasoumi and Ebrahim Bagheri

Department of Electrical and Computer Engineering, Ryerson University, Canada

Laboratory for Systems, Software and Semantics (LS3)

† a.pourmasoumi@ryerson.ca, bagheri@ryerson.ca

One of the most valuable assets of an organization is its organizational data. The analysis and mining of this potential hidden treasure can lead to much added-value for the organization. Process mining is an emerging area that can be useful in helping organizations understand the status quo, check for compliance and plan for improving their processes. The aim of process mining is to extract knowledge from event logs of today's organizational information systems. Process mining includes three main types: discovering process models from event logs, conformance checking and organizational mining. In this paper, we briefly introduce process mining and review some of its most important techniques. Also, we investigate some of the applications of process mining in industry and present some of the most important challenges that are faced in this area.

## 1. Introduction

Most organizations spend a lot of resources for implementing, analyzing and managing their business process models. Hence, tools or techniques that can help managers reach these goals are desirable. Process mining is a new research agenda, which helps managers gain more insight about their organization's processes. The main goal of process mining is to extract process-centric knowledge from event logs of existing information system of organizations. Process mining can be considered to be the x-ray machine, which shows the reality that occurs within the organization. In many cases, the process that is executed in an organization can have many differences with the process that is expected to be running. This can be because of several reasons such as management changes, infractions and so on. Process mining extracts valuable knowledge for managers and brings transparency for them by analyzing event logs that are stored in the database of information systems of organizations.

Process mining is a bridge between data mining and process modeling/analysis. Process mining and data mining have many commonalities. Most of data mining techniques such as classification, clustering, and sequence mining can be used in process mining as well. For example, extracting the bottlenecks of a process, improving a process, detecting the deviations in a process, analyzing the performance of a process, identifying the best and worst employee involved in a process are the types of objectives that require data mining techniques applied on business processes. The main difference is that data mining is data-oriented while process mining is process-oriented. Process mining techniques include three main classes: *i*) process discovery (automated discovery of process model from event log), *ii*) conformance checking (detecting deviations by comparing process model and the corresponding event log), *iii*) organizational mining (including several techniques such as social network analysis, prediction and recommendation systems).

Process mining techniques often work with event logs as input. Each event data to be usable should have at least three properties: *i*) data should have timestamps, *ii*) activity labels should be present and iii) case id of each record should be specified (case id is the id of each process instance). Therefore, there is need to standardize the logging format of event logs. The "*IEEE Task Force on Process Mining*" has developed an xml-based XES[*] logging format. The XES is supported by OpenXES library and many tools such as ProM, Disco, and XESame.

In 2012, the "IEEE Task Force on process mining" also presented a manifesto for process mining, which is translated to more than ten languages [1]. In this manifesto, process mining is introduced and the most challenges and future works of this domain is specified.

### 1.1. *Quality measures in process mining*

Before discussing any process mining approach, it is better to talk about the quality measures in process mining. The most useful quality criteria are as follows [1]:

(a) Fitness: this criteria indicates how much a discovered process model is in accordance with a corresponding event log. A common way for calculating fitness is replaying all traces on the process model and calculating the number of cases that trace cannot be replayed on the process model. The process model that has too much fitness, might suffer from the *overfitting* problem.

(b) Simplicity: the more the discovered model is simple, the more desirable it would be. Various ways of computing simplicity based on the number of activities and relations can be found in [2].

(c) Precision: a model is precise if does not allow for too much unobserved behavior in event logs. For example, in "flower model", all traces of any event log set (which have the same set of activities) can be replayed (Figure 1). Such models can lack precision and hence suffer from an *underfitting* problem.

---

[*] - www.xes-standard.org

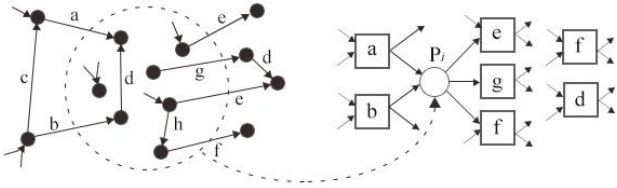

Fig. 5. Mapping of a region into a place

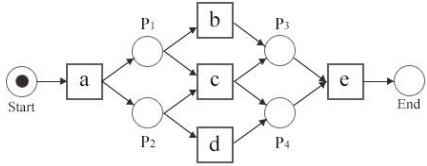

Fig. 3. Discovered process model for event logs
$L_1 = [\langle a,b,d,e \rangle^2, \langle a,d,b,e \rangle^2, \langle a,c,e \rangle^3]$

(d) Generalization: this criteria is used for avoiding the overfitting problem. A model should have a minimum generalization and not be restricted to the behavior that seen in the logs. Because we should always consider that the event logs might be incomplete.

In many cases, these criteria are competing with each other and increasing one may decrease another (e.g., a model that has high fitness might have low generalization). So, process mining algorithms should balance between these quality dimensions.

## 2. What is Process Mining?

### 2.1. *Process discovery*

The first main type and one of the most challenging tasks of process mining is process discovery. A process discovery technique takes an event log of an information system as input and generates a model without using any a-priori information [2]. Process discovery can be investigated from various perspectives, e.g., the control-flow perspective, the organizational perspective, the case perspective and the time perspective. The control-flow perspective focuses on the control-flow of process models such as extracting activity orders in terms of a modeling language (e.g., BPMN [3], Petri net [4], EPCs [5] and UML activity diagram [6]). The organizational or resource perspective focuses on organizational resource (e.g., human, monetary, raw materials and capital) information that can be extracted from event logs. In the case view, we attend to various properties of cases (e.g., count of cases). The time view is related with timing and frequency of events. Process discovery through each of these perspectives can give different valuable insight

|   | *A* | *b* | *c* | *d* | *e* |
|---|---|---|---|---|---|
| *a* | # | → | → | → | # |
| *b* | ← | # | # | ∥ | → |
| *c* | ← | # | # | # | → |
| *d* | ← | ∥ | # | # | → |
| *e* | # | ← | ← | ← | # |

Fig. 2. Footprint matrix of $L_1$

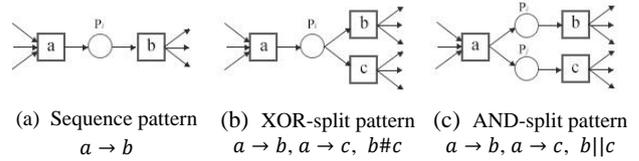

(a) Sequence pattern
$a \to b$

(b) XOR-split pattern
$a \to b, a \to c, b\#c$

(c) AND-split pattern
$a \to b, a \to c, b\|c$

Fig. 4. Some of the most widely used

to managers. In the following, we introduce a classification of process discovery approaches.

#### 2.1.1. *Region-based process mining*

The Alpha algorithm is one of the simplest and practical discovery approaches that produces a Petri net explaining the process behavior recorded in event logs [7]. The idea of the Alpha algorithm is simple and used by many process mining algorithms. The α-algorithm scans the event log for particular patterns. For example, if in the event logs, activity *a* is always followed by *b* but *b* is never followed by *a*, then it is assumed that there is a causal dependency between *a* and *b* [8]. There are four relations in α-algorithm that we briefly explained it:

- Direct succession: $a > b$ if for some case *a* is directly followed by *b*,
- Causality: $a \to b$ if $a > b$ and not $b > a$,
- Parallel: $a \| b$ if $a > b$ and $b > a$,
- Choice: $a \# b$ if not $a > b$ and not $b > a$

Event logs should be scanned for extracting these relations. The result can be shown as footprint matrix as shown in Figure 2. Suppose that $L_1$ is a simple log describing the history of seven cases:

$$L_1 = [\langle a,b,d,e \rangle^2, \langle a,d,b,e \rangle^2, \langle a,c,e \rangle^3]$$

Each trace in $L_1$ corresponds to a possible execution in Petri net process model of Figure 3. Scanning $L_1$ led to footprint matrix of Figure 2. Using the footprint matrix, the particular patterns as shown in Figure 4 can be discovered easily. The detailed algorithm can be found in [8].

Although α-algorithm can discover a large class of process models, there are several limitations. In the face of one-length loop, two-length loop, invisible tasks, duplicated tasks, and non-free choice constructs, the α-algorithm cannot work properly. Hence, several extensions of the α-algorithm like as $\alpha^+$-algorithm [9], $\alpha^{++}$-algorithm [10] is proposed to overcome these problems. However, for real-life event data more advanced algorithms are needed to better balance between different discovery quality measures.

2.1.2. *Region-based process mining*

Region-based process mining approaches are highly based on the *theory of regions* [11]. The theory of regions makes a link between transition systems and Petri nets through the so-called *Net synthesis*. The main idea of theory of regions is that a state-based model such as transition system can be transformed into a Petri net. In [2], several functions for converting event logs to a transition system is introduced. There two main region-based process mining types: a) state-based regions [11], b) language-based regions [12]. In state-based region process discovery, a region is defined as set of states such that all activities in the transition system "agree" on the region [2]. Based on each region, all activities can be classified into entering the region, leaving the region, and non-crossing. After extracting regions based on these simple rules, each minimal region corresponds to a place in Petri net model. In Figure 5 an example region and its corresponding place is shown. In this region, activities *a* and *b* enter to region and activities *e*, *g* and *f* exit the region. Also, activities *d* and *h* do not cross the region. This region can be mapped into a place $P_i$ such as one is show in Figure 5. In similar vein, all regions extracted and a corresponding place for each of them is created.

The second main type of region based process mining is language-based [12]. Similar to state-based algorithms, the aim of this type is to determine Petri net places; however, these models use some specific pre-defined "*language*" instead of a transition system as input. The basic idea of language-based approaches is based on the fact that removing place $P_i$ will not remove any behavior, but adding place $P_i$ may remove some possible behaviors in the Petri net [2]. So, adding a place to a Petri net may restrict the corresponding behavior in the event log. Based on this simple idea, language-based approaches try to add various places as long as the behaviors seen in the logs are not limited. This problem can be modeled in terms of an inequation system. The main obstacle of this category is that linear inequation systems have many solutions and they need to consider the log to be complete. There are some newer approaches for solving these problems such as [13].

2.1.3. *Heuristic mining*

Heuristic mining algorithms use *casual nets* representation. A causal net is a graph whose nodes are activities and arcs are causal dependencies. Each activity has a set of possible input bindings and a set of possible output bindings. In heuristic mining, the frequencies of events and sequences is considered [14] [15]. In fact, heuristic mining aims to remove infrequent paths. For this purpose, at first, similar to footprint matrix in the α-algorithm, a dependency matrix is calculated. A dependency matrix, shows the frequency of *directly follows* relation (i.e., |a > b| is the number of times that *a* is directly followed by *b*) in the event logs. Using the dependency matrix, the so-called *dependency graph* is

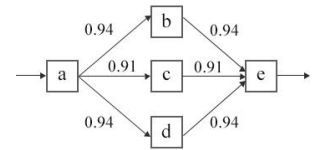

Fig. 6. Dependency graph

created (Figure 6). The dependency graph shows the arcs that meet certain *thresholds*. The main drawback of this presentation approach is that it cannot show the routing logic of business process correctly. For example, in Figure 6, the exact relation type between b, c and d is not clear. However it can show the main stream of process [14]. After extracting dependency graph, it should be converted to casual nets. The detailed algorithm can be found in [14] [15].

2.1.4. *Evolutionary process mining*

Evolutionary process mining is a subcategory of the larger category of search based process mining [16]. Similar to other applications of evolutionary approaches in other domains, evolutionary process mining techniques use iterative procedure, are non-deterministic and have four main steps [2]:

(i) Initialization: in this step the initial population including a set of individual process models is generated randomly. These process models are generated using activities that exist in the event logs, however they might have little compliance with the seen behavior in the event logs, due to the random generation procedure.

(ii) Selection: in this step, the best generated individuals should be selected based on a fitness function for mutation (regenerate step). The fitness function calculates the quality of the discovered process according to existing event logs based on process discovery quality criteria [1]. The best individuals are selected based on the highest fitness score.

(iii) Regenerate: after selecting the best individuals, new generation of individuals is created using two operators: *crossover* and *mutation*. Using crossover operator (the red dotted line in Figure 7), individuals are separated into children. Next, these children are modified using mutation to generate the next generation. The new generation would be considered as the initial population and the algorithm iterates as long as the termination constraint is not satisfied.

(iv) Termination: the algorithm will terminate when the newly generated individuals reach a minimum fitness score. These individuals will be returned as the discovered process model.

In Figure 7, an example of procedure of genetic process mining is shown. There are two individual process models which using *crossover*, each of them are splitted into two sub-processes. Then, using *mutation* the first child of the first process model is composed with the second child of the second process model and second child of the first process is composed with the first child of the second process.
The main drawback of genetic process mining approaches is their time complexity and their uncertainty [16].

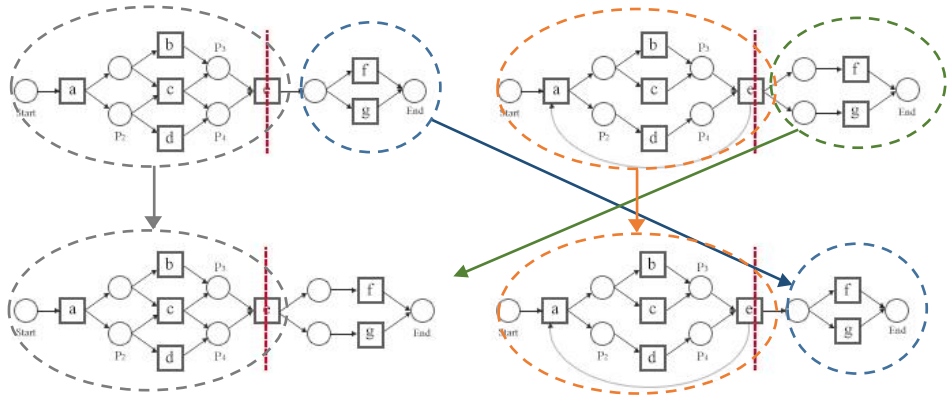

Fig. 6. An example of genetic process mining

## 2.2. *Conformance checking*

The second major type of process mining is conformance checking [2]. Conformance checking is used for deviation detection, prediction, decision making and recommendation systems. In conformance checking, an event log is compared with its existing corresponding process model and it reveals that if process model conforms to reality and vice versa. For example, in educational systems, using conformance checking it can check which students do not select their courses according to their educational chart and have deviations. As another example, in the banking domain, it can check which cases do not follow the process model and why it happens? Scanning event logs using conformance checking techniques increases transparency and leads to fraud detection [17] and is very useful tool for auditors. Conformance checking also used for evaluating process discovery approaches from a *fitness* point of view.
One of the most widely used methods for conformance checking is replaying all cases of event log using a token on its corresponding process model [17] [18]. Based on this method, the *fitness* of the event log in light of the process model is calculated (the fitness is most relevant measure from the four quality criteria for calculating conformance). In [2], fitness is defined as "the proportion of behavior in the event log possible according to the model". Token-replay based techniques are simple and can be implemented efficiently [2].
The main drawback of token-replay based methods is that those cases that do not fit the process model would be ignored totally. So, in this situation, the result might be biased and diagnostics would be decreased. Furthermore, token-replay based methods are only Petri-net specific. There are another class of methods for conformance checking such as *alignment*-based techniques [2]. Alignment-based approaches have been developed to overcome these problems. In these approaches, a table mapping between each trace and process model is created [19]. These mappings are called alignment or optimal alignment. Using alignment-based approaches, each case can be analyzed separately and the results can be aggregated at the process model level [2]. More details on alignment-based approaches can be found in [20] [21] [19].

## 2.3. *Organizational mining*

Organizational mining is the third type of process mining activity, which brings more insight for organizations and can led to added value. Using organizational mining, the bottlenecks of processes is analyzed and new improvements are proposed. The most widely used technique for this purpose is social network analysis [8]. The resources are represented by nodes and the relations by links. The thickness of a link shows the amount of relationships between two nodes [8]. Also, the nodes or links may have weight which show their importance. So far, many metrics for analyzing social networks have been proposed such as *closeness, betweenness, centrality*, and *shortest distance* among others [2].
Another application of organizational mining is extracting organizational structures or organizational charts. In reality, in many cases the discovered organizational chart from event logs differ from existing organizational structure on the documents. Based on the discovered chart, the senior manager can make decisions, e.g., change the physical location of the employee based on their real connections. For more details on organizational mining see [22].
One main aspect of organizational mining is cross-organizational mining [1]. In cross-organizational mining more than one organization is involved. Actually, the cross-organizations collaborations is usually in two modes. First, different organizations work with each other on the same instance of a process. For example, for building a house several organization such as municipality, insurance, power authority and others might be involved. In this case, the organizations act like puzzle pieces [1]. In the second type, several organizations share a common infrastructure and execute the same process model. The example of this case is cloud based software services. A famous example is *Salesforce.com,* which manages and supports the sales of processes for many organizations [2].

## 3. Applications of Process Mining

Process mining has been used in several domains such as healthcare, financial (specially banking domain), production, e-commerce, logistics, monitoring, e-government, insurance, among others. Here, we have presented two important applications of process mining in today's life: healthcare and e-commerce. In the following, we give an example of application of process mining in these domains.

### 3.1. *Process mining in healthcare*

One of the most important challenges in healthcare is rising costs of healthcare. Based on OECD health statistics 2015, average per capita health spending across OECD countries has been on an upward trend†. So, there is urgent need for decreasing these costs. One ideal solution is focusing on the complex time-consuming treatment processes such as cancer or surgery treatment processes [24] [25]. Traditional approaches for improving and redesigning such process models is conducting interviews and field studies. Analyzing process models in this way is too costly and time consuming. Moreover, depending on the person being interviewed and depending on the interviewer, it can be subjective. Furthermore, in some cases, there might be organizational resistances and many stakeholders may not tend to decrease costs or have any changes in the process models. So in this case, we need to capture the reality from existing information systems. Process mining tools can give objective suggestions for reducing the treatment processing time of patients.

For further information, you can refer to [26]. Man et al. published an interesting book about applications of process mining in the healthcare domain and the challenges that exist in this regards [26].

### 3.2. *Process mining in e-commerce*

E-commerce has significant portion of business in today's world. The most worthy thing in e-commerce is data. As more data is available, further analysis can be done and more and more customers will be attracted. For example, Amazon as one of the biggest marketplace seller uses process mining for analyzing the online purchase process of customers. Combining process mining algorithms with natural language processing methods such as sentiment analysis can give clear insight into each user's behavior and led to accurate recommendations. Suppose a customer who wants to purchase a camera. Certainly, the customer investigates and compares various brands and models to make the most optimal decision and finally decide to purchase a camera of model X. In the meantime, the customer would read other user's comments (feedbacks). Having clicking/scrolling data of different customers that purchased camera of model X, using process discovery techniques the seller can discover the path that most of the customers follow to purchase camera of model X. Based on the discovered path, the seller can make some decisions to change this path to decrease the time to decision of customers and also selling more camera of model X.

### 3.3. *Other applications of process mining*

Process mining has also seen widely increasing applications in banking/financial domain [27]. Todays, large number of banks and financial institutes use process mining techniques in threefolds: *i*) improving their own inter-organizational processes, *ii*) in order to increase income and attract more customers, *iii*) auditing financial accounts [28].

Process mining has also been used successfully in the insurance domain. In [29], a case study of applications of process mining in one of the largest insurance companies in Australia (Suncorp) is reported. This paper reported new insight of the way insurance claims has being processed at Suncorp.

Process mining also has been used in many *municipalities* [30], *hardware manufacturers* like as ASML [31], *educational* organizations such as universities, (e.g., TU/e employed process mining to analyze the educational behaviors of the students), *transportation* industry, *cloud computing* based industry such as Salesforce.com and others.

## 4. Challenges

Despite the many capabilities and applications of process mining techniques, there are still important challenges that should be addressed [1] [2]. The most important challenge of process mining is data. There are several concerns about event data; *a*) data usually are *object-oriented*, not *process-oriented* (there is need for non-trivial efforts to convert object-oriented event data to process-oriented data); *b*) event data might be *incomplete*; *c*) data might be distributed over several sources with different terminology, *d*) in most of situations there is need to do data cleansing on processes, *e*) different levels of *timestamps* ranging from milliseconds to coarse format e.g., 23-08-2016 might exist, and *f*) the activities names may differ in different data sources.

Another important concern about processes is *concept drift* [32]. Here, concept drift means that a process might change during analysis. There are several reasons that can lead to these changes. The data can change because of business conditions. For example, in some periods of the year like such as Christmas time or before the opening of schools, store sales change considerably, or in some days such as Sundays there is less need for the presence of employees. Hence it is important to select appropriate range of data. Sometimes, event data changes due to the changes in process models or event in information systems. For example, a new manager may want to make some changes in running processes. So, there will be a concept drift between data related to old process and the newer process. In [33], four types of concept drift have been introduced: sudden drifts, gradual drifts, recurring drifts and incremental drifts. In a sudden drift, the whole process will suddenly change from start to end (e.g., due to management changes the whole process changes). In gradual drift, the process model changes for the non-uniform interval of time. For example, according to requests from different stockholders or even the employee, the process changes gradually over time. Recurring drifts refers to cyclical changes that caused by business conditions (e.g., as changes in store sales in certain period of time). In incremental drift, a process will change in a gradual fashion. Considering concept drift is important issue in dealing with real processes logs. In the recent years, several papers have been published in this domain [32] [34] [35]. In [36] a review on concept drift in process mining is presented.

---

† - https://www.oecd.org

The different level of granularity of event logs is another important issue that should be considered. For example, in the health care domain, hospital information systems (HIS) cover wide range of treatment processes; from the simple outpatient to complex surgical or cancer treatment procedures [26].

**5. Process Mining Tools**

There are several free open source and commercial software for process mining. ProM[‡] is the most widely used and complete open-source process mining platform [37]. ProM is multi-perspective and includes several modules for process discovery, conformance checking and organizational mining. Most of the process discovery and conformance checking algorithms is implemented in ProM and can be accessed freely. There are also some other non-commercial process mining tools such as RapidProM[§](combination of RapidMiner with some process mining plug-ins from ProM) [38], PMLAB[**](a script-based process mining tool which support variety of process discovery techniques, specilly region-based approaches), CoBeFra[††](a framework for conformance checking), PLG[‡‡] (a platform for generating random process models and corresponding event logs) [38] and PVLG (a tool for generating process variants) [39].

There are also several commercial process mining software such as Disco[§§], Celonis[***], Minit[†††], myInvenio[‡‡‡], QPR Process Analyzer[§§§], Rialto[****], SNP[††††], Interstage Business Process Manager Analytics (Fujitsu) etc. In [2], a comprehensive classification of these products based on the different criteria (e.g., openness, formalness of discovered process, input types, output types, and types of deployments) is presented. Moreover, an analysis of strengths and weaknesses of these tools is presented.

**5.1. *Conclusion & Future directions***

We have briefly review the state of the art in process mining. We introduced the main works in process mining and presented well known approaches and tools. Furthermore, we introduced some the open problems and challenges in this field. Handling incomplete, noisy, object-oriented event data in proposing different process mining algorithms should be considered.

With increasing amount of data in today's databases, there is need for techniques that can handle such big data. There are some works on distributed systems for handling and processing event logs [40] [41]. However, there is yet need for more work for developing process mining algorithms based on big data infrastructures

Online process mining is another research direction that allows for the real time processing of event and process data. Along with the increase in the quality of process mining techniques, the time complexity of these algorithms should also be considered.

Analyzing cross-organizational mining is also an attractive research area. In cross-organizational mining, there are more challenges compared to single organizational mining, because each organization has its own organizational culture and its own process structure and infrastructure. Mining process models through the event logs of different information systems can be challenging.

---

[‡] - http://www.promtools.org
[§] - www.rapidprom.org
[**] - https://www.cs.upc.edu/~jcarmona/PMLAB/
[††] - http://www.processmining.be/cobefra.
[‡‡] - http://plg.processmining.it/
[§§] - www.fluxicon.com

[***] - www.celonis.de
[†††] - http://www.minitlabs.com
[‡‡‡] - https://www.my-invenio.com
[§§§] - http://www.qpr.com/products/qpr-processanalyzer
[****] - www.exeura.eu
[††††] - www.snp-bpa.com